\documentclass[PRL,superscriptaddress,reprint]{revtex4-1}

\usepackage{graphicx}
\usepackage{dcolumn}
\usepackage{bm}
\usepackage{graphicx}
\usepackage{dcolumn}
\usepackage{bm}
\usepackage[utf8]{inputenc}
\usepackage{mathptmx}
\usepackage{fancyhdr}
\usepackage{amsmath,amsthm,amssymb}
\usepackage{graphicx}
\usepackage{hyperref}
\usepackage{lipsum}
\usepackage{textcomp}
\usepackage{dsfont}
\usepackage{tabularx}
\usepackage{tikz}
\usepackage{physics}
\usetikzlibrary{scopes,calc,arrows}
\usepackage[makeroom]{cancel}
\usepackage{enumitem}
\usepackage{listings}
\usepackage{color}
\usepackage{slashed}

\begin{document}
	
	\preprint{APS/123-QED}
	
	\title{Experimental Evidence of Plasmoids in High-$\beta$ Magnetic Reconnection}

	\author{J. A. Pearcy}
	\email{pearcy@mit.edu}
	\affiliation{%
		Plasma Science and Fusion Center, Massachusetts Institute of Technology, Cambridge MA 02139 USA
	}%
	\author{M. J. Rosenberg}
	\affiliation{Laboratory for Laser Energetics, University of Rochester, Rochester NY 14623
	}%
	\author{T. M. Johnson}%
	\author{G. D. Sutcliffe}%
	\author{B. L. Reichelt}%
	\author{J. D. Hare}%
	\author{N. F. Loureiro}%
	\author{R. D. Petrasso}%
	\author{C. K. Li}%
	\email{ckli@mit.edu}
	\affiliation{%
		Plasma Science and Fusion Center, Massachusetts Institute of Technology, Cambridge MA 02139 USA
	}%
	
	\date{\today}
	
	\begin{abstract}
		Magnetic reconnection is a ubiquitous and fundamental process in plasmas by which magnetic fields change their topology and release magnetic energy. Despite decades of research, the physics governing the reconnection process in many parameter regimes remains controversial. Contemporary reconnection theories predict that long, narrow current sheets are susceptible to the tearing instability and split into isolated magnetic islands (or plasmoids), resulting in an enhanced reconnection rate.  While several experimental observations of plasmoids in the regime of low- to intermediate-$\beta$ (where $\beta$ is the ratio of plasma thermal pressure to magnetic pressure) have been made, there is a relative lack of experimental evidence for plasmoids in the high-$\beta$ reconnection environments which are typical in many space and astrophysical contexts. Here, we report the observation of strong experimental evidence for plasmoid formation and dynamics in laser-driven high-$\beta$ reconnection experiments.
	\end{abstract}
	
	\maketitle

	Magnetic reconnection is the process by which magnetic fields in plasmas change their topologies and release magnetic energy \cite{cit1,cit2}. It is a phenomenon with widespread importance to many fields of physics, from astrophysics \cite{cit1,cit2,cit3,cit4,cit5} to laboratory and fundamental plasma physics \cite{cit1,cit2,cit3,cit6,cit7,cit8,cit9,cit10,cit11}. The theoretical understanding of magnetic reconnection has evolved significantly over the history of plasma physics. The classical Sweet-Parker model of reconnection \cite{cit12,cit13} uses dimensional arguments to infer parameters such as the width of a long, thin, steady-state current sheet. Its fundamental prediction is that the current sheet width is $\delta_\text{SP}=LS^{-1/2}$, where $L$ is its length and $S=\mu_0Lv_A/\eta$ is the Lundquist number \cite{cit12,cit13}, with $v_A$ the Alfv\'en velocity computed with the reconnecting field and $\eta$ is the plasma resistivity. Consequently, the Sweet-Parker reconnection timescale is $\tau_\text{SP}\sim S^{1/2}L/v_A$. Since typical reconnecting plasmas have $S\gg1$ as a result of very small resistivity, the Sweet-Parker timescale is orders of magnitude too large to explain observations of reconnection in astrophysical and laboratory contexts.
	
	Modern reconnection theories and associated simulations \cite{cit14,cit15,cit16,cit17,cit18,cit19,cit20} have revealed that the long, thin current sheets predicted by the Sweet-Parker model are vulnerable to the fast-growing tearing instability. This instability is predicted to lead to the formation of chains of isolated magnetic islands, known as ``plasmoids”, which enhance the reconnection rate and associated dissipation of magnetic energy by eliminating the dependence of the reconnection rate on the global Lundquist number of the current sheet \cite{cit17,cit20}. Plasmoids are thought to be a generic feature of large-scale reconnecting systems, having been observed in a wide range of parameter regimes.
	
	\begin{figure}[h!]
		\centering
		\includegraphics[scale=0.3]{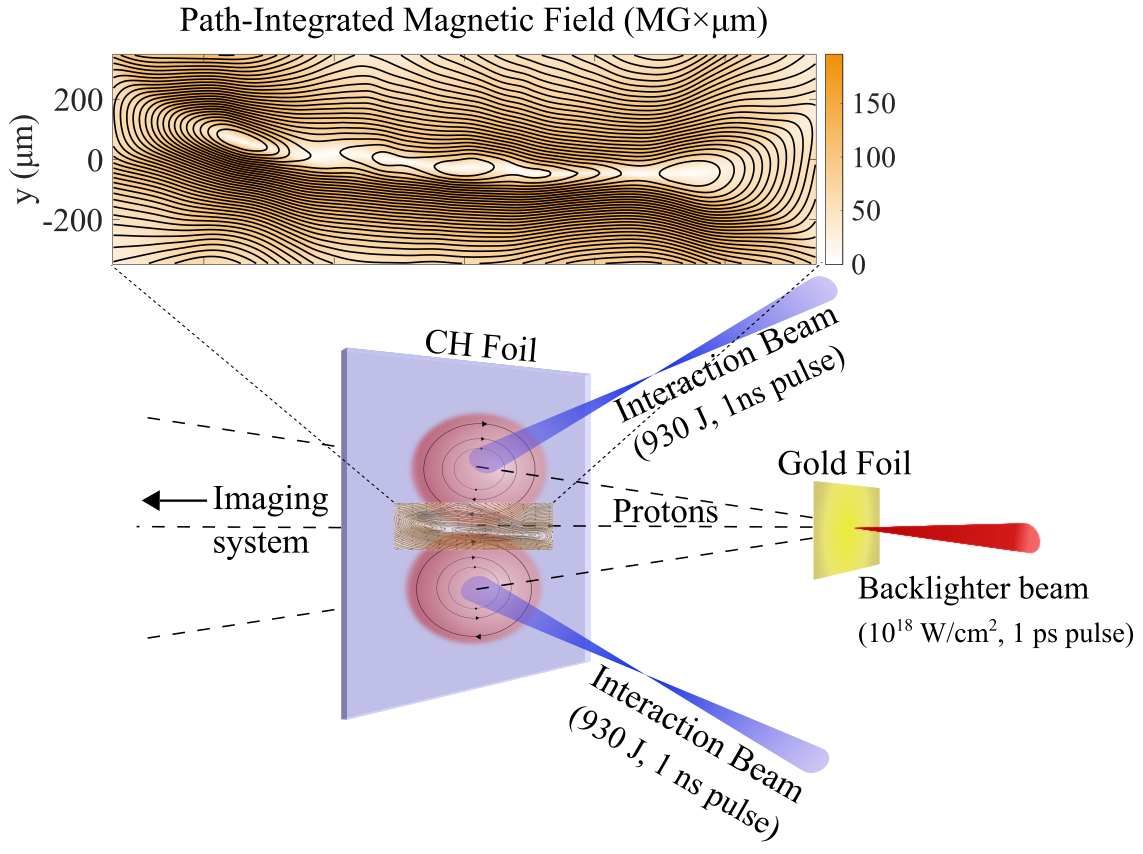}
		\caption{\textbf{Experimental design.} A plastic (CH) foil illuminated by two interaction laser beams, producing plasma bubbles. A proton backlighter consisting of a gold foil driven by a high-power, short-pulse laser beam produces energetic protons, producing proton images. The collision between two antiparallel magnetic fields driven by the bubble expansion leads to the changing of the field topology and reconnection, as indicated in the reconstructed fields shown in the insert.}
		\label{expdesign}
	\end{figure}
	
	To date, most magnetic reconnection experiments performed to seek out observations of plasmoids have investigated two regimes: relatively tenuous quasi-steady-state magnetically driven plasmas for which the typical plasma $\beta$ (the ratio of thermal pressure to magnetic pressure) is $\beta\ll 1$ \cite{cit7,cit8}; and pulsed-power driven plasmas for which $\beta\lesssim1$ \cite{cit9,cit21}. In magnetically driven experiments, plasma inflows remain sub-Alfv\'enic; in pulsed-power driven systems, super-Alfv\'enic flows and associated flux pileup were observed \cite{cit21}.
	
	\begin{figure}[h!]
		\centering
		\includegraphics[scale=0.32]{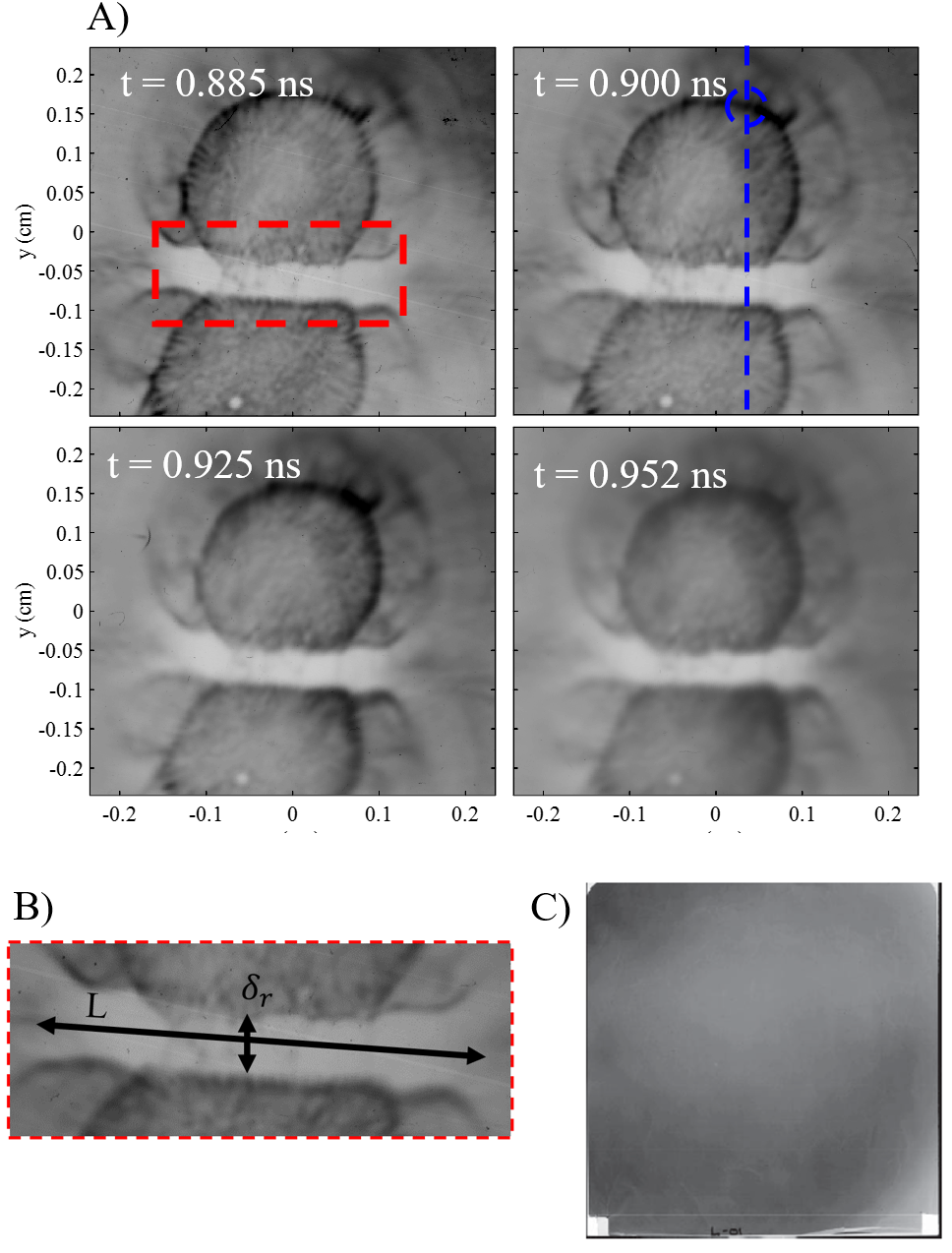}
		\caption{\textbf{Raw radiography data. }A) Proton radiographs show the spatial structure and temporal evolution of magnetic fields associated with the expanding plasma bubbles and interaction regions. These radiographs were formed by protons with energies between 11 and 36 MeV. The dashed blue line in the second image represents approximately where lineouts for Figure \ref{times}A and \ref{times}B were taken, though each image had a slightly different location for the lineout to ensure sampling of the correct current sheet structure (between apparent plasmoids). The dashed blue circle denotes the position of edge of the Biermann bubble along the lineout. In these images, image contrast has been artificially increased to render the structure in the center of the radiograph more visible to the naked eye. B) Zoomed-in view of the reconnection layer (corresponding region in the full data set is denoted by the dashed red box). The apparent (gross) length and width of the reconnection layer as inferred from the radiograph ($L$ and $\delta_r$, respectively) are labelled. C) A typical null shot (without subject plasma), taken from another experimental campaign, is shown to illustrate what one expects the initial proton flux to look like.}
		\label{data}
	\end{figure}

	We report here the direct observation of plasmoids in a laser-driven reconnection experiment with high plasma $\beta$ ($\sim10$) and super-Alfv\'enic inflows, using advanced proton radiography and deflection-field reconstruction techniques \cite{cit23,cit24} which allow unprecedented insight into the structure of magnetic fields in HEDP experiments, alongside time-resolved Thomson scattering to characterize important plasma parameters \cite{cit25}. While previous laser-driven experiments have been performed to investigate magnetic reconnection (ex., \cite{cit22}), prior campaigns have not directly observed plasmoid formation; in contrast our results show strong direct evidence of plasmoids. The conditions in our experiment manifest experimental regimes which are typical of laser-produced plasmas in laboratory experiments \cite{cit10,cit11,cit26}, and which match certain parameters of astrophysical plasmas, such as the high-$\beta$ plasmas of the intra-cluster medium \cite{cit27,cit28} or the Galactic center \cite{cit29}.
	
	Our laser-produced plasma magnetic reconnection experiment was carried out at the OMEGA EP laser facility \cite{cit30} in 2012. Initial findings on a subset of the data  were reported by Rosenberg et al \cite{cit33}. In this study, we use modern, sophisticated analysis techniques to investigate the remainder of the large data set from the shot day. In addition, to characterize plasma parameters an experiment with nominally identical drive conditions was carried out at the OMEGA Laser Facility in 2019 to perform Thomson scattering measurements.
	
	The experimental setup is shown schematically in Figure 1. In the experiment, the subject target is a 12 $\mu$m thick plastic (CH) foil, driven by two 1-ns square pulse interaction beams of 930 J each, with spot sizes of $800\ \mu$m and a separation of 1400 $\mu$m. The interaction beams impinging on the CH foil produced two hemispherical plasma bubbles. The Thomson scattering measurements indicate that such bubbles have a typical plasma temperature $T_e\sim2\ $keV, electron density $n_e\sim3\times10^{19}\ $cm$^{-3}$, and bubble expansion velocity $v\sim500-800\ \mu$m/ns immediately prior to collision (see Table \ref{table}); the results of this analysis are roughly consistent with DRACO simulations of our experiment \cite{cit33}.
	
	The bubbles produced circulating magnetic fields via the Biermann battery mechanism \cite{cit31}, wherein a magnetic field is generated due to misaligned temperature and density gradients: $\partial_t\vec{B}\propto\nabla T_e\times\nabla n_e$. As illustrated in Figure \ref{expdesign}, the two plasma bubbles expand into each other, compressing their antiparallel magnetic fields and driving the reconnection process.
	
	\begin{table}[b]
		\caption{\label{table}%
			Quantities above the horizontal line are measured from our radiography and Thomson scattering data; quantities below the horizontal line are calculated based on those measurements. The “Magnetic Field” quantity is obtained by assuming a path-length traversed by backlighting protons and calculating based on the reconstructed path-integrated values
		}
		\begin{ruledtabular}
			\begin{tabular}{cc}
				\textrm{Parameter}&
				\textrm{Approximate Value}\\
				\colrule
				Plasma density, $n_e$ & $3.2\times10^{19}\ $cm$^{-3}$\\
				Electron temperature, $T_e$ & $ 2.2$ keV \\
				Magnetic field (near CS), $B$ & $40-70$ T \\
				Length of current sheet, $L$ & $\sim2000$ $\mu$m \\
				Width of current sheet, $\delta$ & $20-60$ $\mu$m \\
				\hline
				Plasma beta, $\beta$ & $\sim$10 \\
				Ion skin depth, $d_i$ & $\sim$55 $\mu$m\\
				Electron skin depth, $d_e$ & $\sim1$ $\mu$m\\
				Spitzer resistivity, $\eta$ &$\sim$$4\times10^{-2}$ $\Omega\cdot\mu$m \\
				Sound speed, $c_s$ & $\sim$4.69$\times10^{11}$ $\mu$m/s\\ 
				Alfv\'en speed, $v_A$ & $\sim$$1.5\times 10^{11}$ $\mu$m/s \\
				Lundquist number, $S$ & $\sim$$9000-15000$ \\
				CS width/ion scale,$\delta/d_i$ & $\sim0.2-1$ \\
				Sweet-Parker width, $\delta_{SP}$ & $\sim$$15-20$ $\mu$m \\
			\end{tabular}
		\end{ruledtabular}
	\end{table}

	\begin{figure*}
	\includegraphics[scale=0.5]{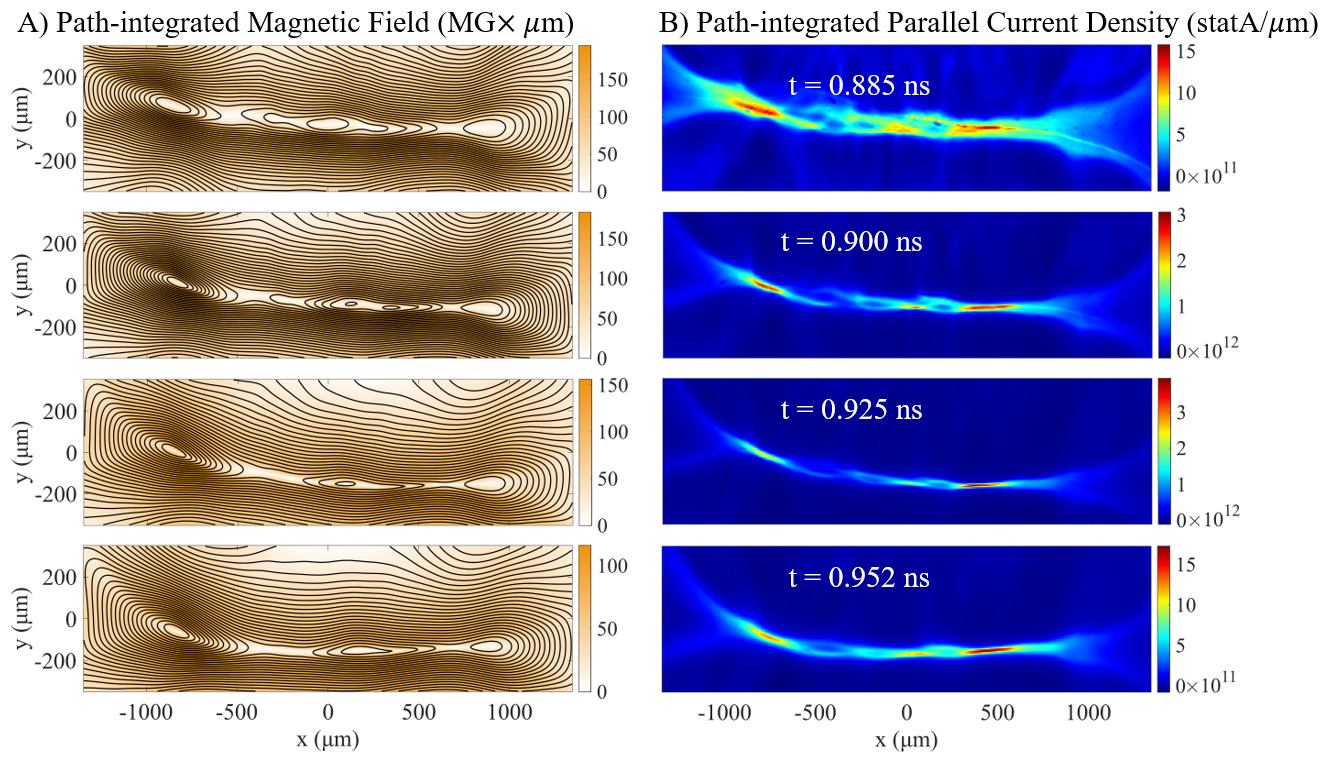}
	\caption{\textbf{Reconstructed path-integrated fields and currents.} A) The reconstructed magnetic fields from the radiography data. All of these reconstructions are of different image plates from the same experimental shot. The black curves in the magnetic field strength plots denote contours of the magnetic vector potential $\vec{A}$. The presence of isolated magnetic islands, or plasmoids, is evident in all of the radiographs.  B) The inferred path-integrated parallel current, obtained from Ampere’s law by taking the curl of the path-integrated magnetic field.}
	\label{recons}
	\end{figure*}

	Figure \ref{data} presents the proton fluence radiographs of the reconnecting plasma bubbles, imaged with $\sim10-40$ MeV protons, generated by high-power short pulse laser based on target-normal-sheath-acceleration (TNSA) mechanism \cite{cit32}, on a stack of radiochromic (RC) film. Filtration effects lead to each piece of film in the stack being sensitive to a different proton energy, and based on the different times of flight, probe the reconnection region at different times, allowing us to directly image the temporal evolution of the reconnection region.
	
	The deflection of backlighting protons within the experimental plasma conveys the spatial structure and temporal evolution of magnetic fields associated with the two colliding plasma bubbles. The concentration of proton fluence in a circular pattern around each bubble is the consequence of inwards proton deflections caused by large scale, azimuthal Biermann fields. Between the two expanding plasma bubbles, the pale central region with a noticeable deficit of proton fluence represents a reconnection layer with a large length $L$ to apparent width $\delta_r$ aspect ratio. These radiographs provide a direct picture of the reconnection layer, where the magnetic fields deflect protons out of the current sheet towards the center of each individual bubble \cite{cit33}.
	
	In our analysis, we assumed that the structure of each radiograph is determined solely by particles carrying energy near the Bragg peak, where sensitivity is highest; as the peak is relatively narrow and TNSA proton spectra are generally exponentially decaying \cite{cit23}, this assumption is unlikely to introduce significant uncertainty in the analysis. Additionally, Figure \ref{data}C shows a sample null radiography shot, with no subject plasma; characterization of the initial proton flux prior to interaction with magnetic fields is important in the reconstruction procedure.  In a typical null shot, we observe low-amplitude large-scale spatial variations of the order of the image size, but no high-frequency non-uniformities that could be confused with physics effects seen in the reconstructions. This suggests that it is reasonable to infer initial proton fluxes by filtering out high-frequency non-uniformities in the flux images while keeping total proton flux constant.
	
	To quantitatively characterize the field distribution associated with the reconnection layer, the measured radiographs are numerically reconstructed with an finite-difference Monge-Ampere solving algorithm \cite{cit24}. In addition to the previously noted inferences made about the proton flux images, such an algorithm also assumes that proton tracks do not cross each other between the source and the detector screen (i.e., that the relationship between initial proton trajectories and their final positions on the screen is injective). Finally, we assume that magnetic fields are the dominant factor causing proton deflections in the face-on radiography, as has been experimentally validated \cite{double_bubble}. Once these assumptions are made, the reconstruction can be carried out.
	
	 Figure \ref{recons}A shows the path-integrated magnetic field strength inferred from the deflection-field reconstruction process in the reconnection region alongside contours of the magnetic vector potential. The reconnection region is approximately delineated by the dashed red box in the second image of Fig. \ref{data}B. The magnetic field strength increases to a peak off-center of the reconnection layer before decreasing and rapidly switching direction as the center is crossed. This observation matches the intuitive form of the magnetic field one would infer based on the radiography data: the large white region with low proton flux implies the presence of strong fields, while the dark regions on either side indicate that the field points in opposite directions on either side of the center (deflecting protons out to both sides).
	
	More notably, Fig. \ref{recons}A clearly reveals the presence of striking, isolated magnetic islands, or plasmoids, in the reconnection layer, which are clearly identifiable by the appearance of closed magnetic field lines (flux contours). The positive identification of these plasmoids is strongly justified when one considers the well-characterized nature of the backlighting proton flux (Fig. \ref{data}C) and high spatial resolution of the proton radiography ($\sim$5 $\mu$m). We are able to well resolve the measured plasmoid structure (for example, the width of a primary plasmoid is $\sim100-200$ $\mu$m, Fig. \ref{recons}A), thus providing strong experimental evidence of plasmoid formation in this high-$\beta$ reconnection event.
	
	Several important features related to the time evolution of the reconnection layer and its constituent plasmoids are visible in Fig. \ref{recons}A.  First, from 0.885 ns to 0.952 ns we see that the leftmost plasmoid appears to continuously grow in both area and width, although as we discuss later this is unlikely to be a reconnection effect. Meanwhile, there are suggestions in the central region of smaller, second-generation (``secondary”) plasmoids with apparent widths above our predicted spatial resolution of $\sim5$ $\mu$m, though we consider this identification marginal.
	
	The inferred path-integrated current density parallel to the direction of proton propagation (calculated directly from the path-integrated fields by using Ampere’s law) is displayed in Fig. \ref{recons}B, from which we measure a current sheet length $L\sim2000$ $\mu$m (indicating that $L/d_i\sim35$, where $d_i=c/\omega_{pi}$ is the ion skin depth). It is clear from these images that the current sheet is not laminar or uniform, and instead is highly dynamic. One way this manifests is in the transition from an initially rather broad current sheet (at $t = 0.884$ ns) to a much narrower current sheet at later times.
	
	Our interpretation of these data is that as the two Biermann bubbles expand into each other, an extended current sheet is formed whose aspect ratio $L/\delta$ increases as a function of time at supersonic rates. This implies that the current sheet becomes progressively more unstable to the tearing instability \cite{cit34,cit35}, resulting in the onset of plasmoid formation while the width of the current sheet $\delta$ remains above the ion kinetic scale $d_i$. Because our plasma is relatively collisional, we think that it is resistivity that is breaking the frozen flux condition, and therefore that the plasmoid formation occurs in a semi-collisional regime \cite{cit36}. In the subsequent evolution of the current sheet, we observe further shrinkage of its width to below the ion skin depth, as shown in Fig. \ref{times}B, due to the continued expansion of the Biermann bubbles ($v\sim800$ $\mu$m/ns, faster than reconnection timescales) over the course of the experiment.
		
	\begin{figure}[h!]
		\centering
		\includegraphics[scale=0.4]{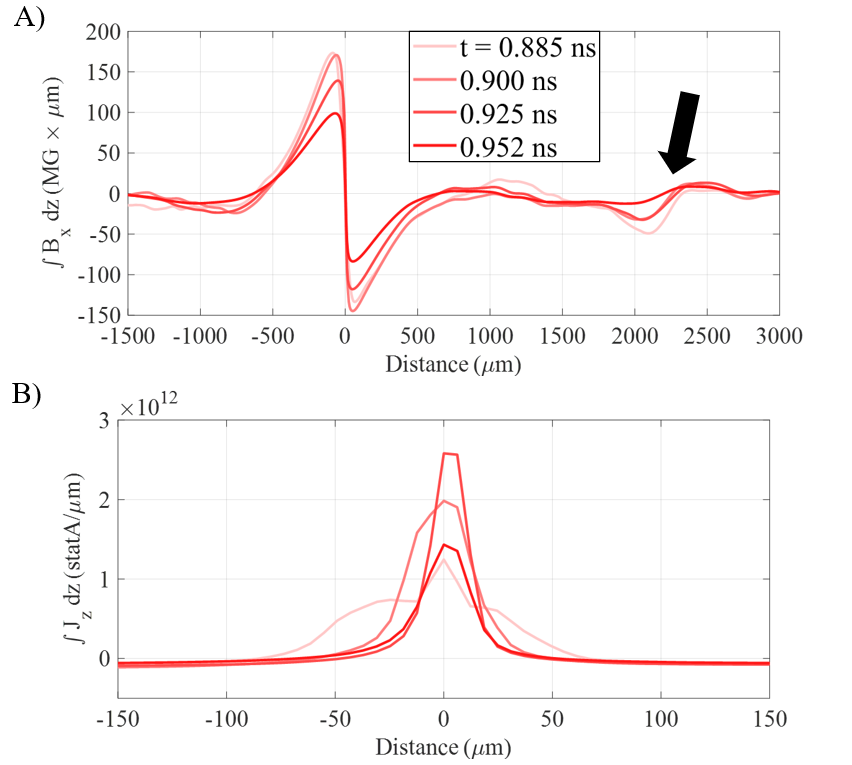}
		\caption{\textbf{Lineouts through the current sheet. }A) Lineouts of the path-integrated $x$-directed magnetic field at each time step, showing the piling up of magnetic flux on either side of the reconnection layer. The black arrow denotes the approximate location of the edge of a Biermann bubble (referred to later as ``far from the current sheet”), demonstrating that the field strength is considerably higher in the reconnection layer than outside of it. B) Lineouts of the path integrated parallel current at each time step (enlarged to show detail – note restricted horizontal length scale).}
		\label{through}
	\end{figure}

	This narrowing effect of the current sheet is explored more fully in Fig. \ref{through}, showing lineouts of the path-integrated magnetic field (\ref{through}A) and lineouts of the path-integrated current density (\ref{through}B) at each time step (lineouts of Fig. \ref{recons}A and 3\ref{recons} respectively). The variations in magnetic field shown in Fig. \ref{through}A give important insight into the reconnection process; the most dramatic observation is an increase in magnetic field strength near the current sheet, when compared with the outer edge of a bubble, providing experimental evidence of the magnetic flux pileup which is expected given super Alfv\'enic inflows. From Fig. \ref{through}B, we can infer the width of the current sheet $\delta$ by measuring the full width at half-maximum (FWHM) of the path-integrated current peaks (shown in Fig. \ref{times}B and discussed below).
	
	\begin{figure}[h!]
		\centering
		\includegraphics[scale=0.41]{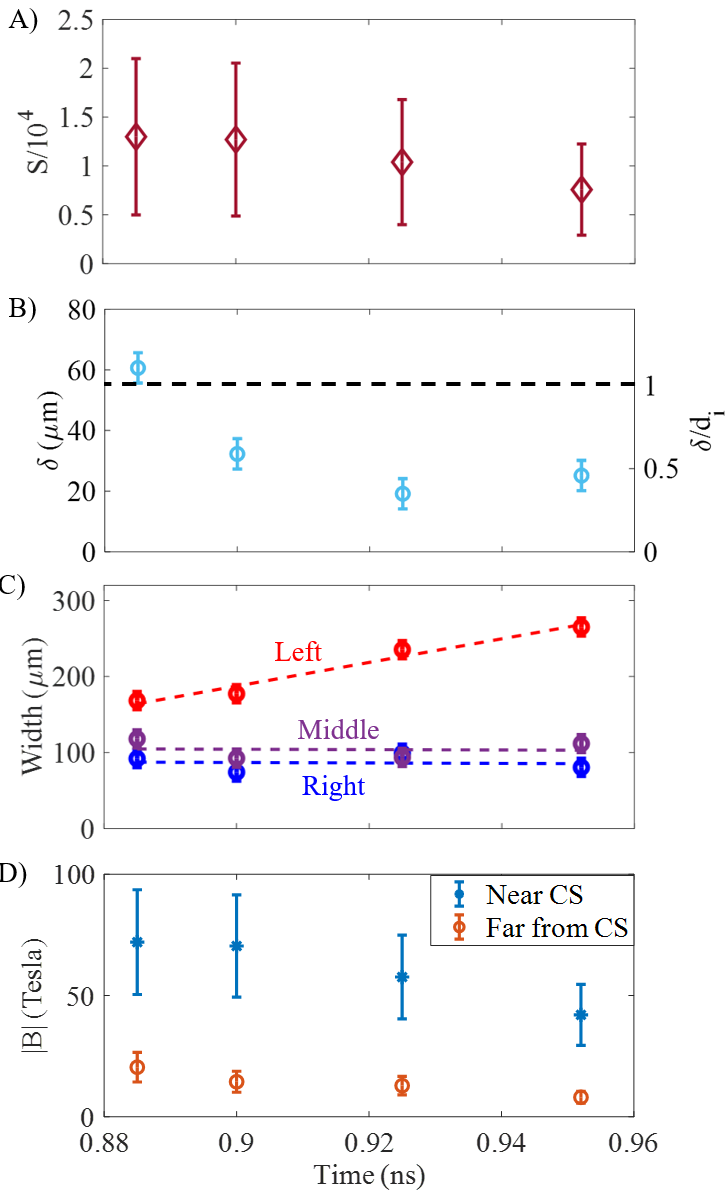}
		\caption{\textbf{Time evolution of various quantities. }A) The temporal evolution of measured Lundquist number is plotted as a function of sampling time.  B) The width of the current sheet, inferred from the lineouts in Fig. \ref{through}, over time. The left axis shows the width in microns, while the right axis shows the ratio of the measured width to the calculated ion skin depth. C) The measured width of the various plasmoids as inferred from the contours of the magnetic vector potential. D) The peak magnetic field near the center of the current sheet plotted alongside the representative magnetic field near the edge of an expanding bubble (“far from CS”, see Figs. \ref{data}A and \ref{through}A).}
		\label{times}
	\end{figure}
		
		Fig. \ref{times}A is the temporal evolution of measured Lundquist number ($S=\mu_0Lv_A/\eta$) as a function of sampling time. Over the course of the experiment, $S$ varies moderately between $\sim9000-13000$; these values are in the highest range of those available to dedicated reconnection experiments, and are large enough that diffusive effects should be minimal in this experiment. Thus transition to a plasmoid-mediated regime can be expected and is indeed observed.
		
		Fig. \ref{times}B displays the time-evolution of the current sheet width through X-points in the current sheet (these are the widths inferred from Fig. \ref{through}B), with the right-hand axis showing the ratio of the width to the ion skin depth. We see that at the earliest time the current sheet is wider than the ion skin depth, but subsequently narrows to below the ion scale as the two plasma bubbles continue to expand into each other.
		
		Shown in Fig. \ref{times}C is the time evolution of the width of the three large plasmoids observed in the experiment. The central and rightmost plasmoids display no increase in width over the experimental timescale, within error bar. This is consistent with theoretical expectations; assuming a nominal reconnection rate of  $\sim0.1 v_A$ (consistent with ion kinetic effects being important), the anticipated increase in plasmoid width is on the order of a few microns, below our experimental resolution. Conversely, the leftmost plasmoid displays an apparent rapid growth that cannot be explained by timescales associated with the reconnection theory. Geometrical effects caused by the finite size of the experimental region could explain this observation; in particular, significant edge effects caused by the flows of the expanding Biermann bubbles could lead to this anomalous growth. Because the bubbles are nearly circular, the region where they collide and interact does not form a long, uniform idealized current sheet and instead has a relatively small region of uniformity flanked by areas where edge effects may be significant. An asymmetry in the flows driving the expansion of one bubble could then contribute to the anomalous growth of the leftmost island. A comparison of the apparent width growth rate ($\sim 1300\ \mu$m/ns) to the bubble expansion velocity ($\sim800\ \mu$m/ns from each side) suggests that this explanation is reasonable.
		
		Meanwhile, Fig. \ref{times}D measures the magnitude flux pileup effects over the timescale of the experiment. We observe that the strength of the magnetic field in a region on the far sides of the expanding plasma bubbles is consistently lower than the strength of the magnetic field in the interior region of the current sheet by a factor which varies between 3 and 6.
		
		We wish to note here that in these discussions the ions have been simply treated approximately as a single species by averaging the hydrogen and carbon (in a 1:1 ratio, and fully ionized). Quantitatively addressing the details of spatial structure and dynamics and species separation associated with a two-ion fluid plasma is a significant challenge for future work.
		
		This work represents a significant step forward in the understanding of high-$\beta$ magnetic reconnection. One aspect we wish to emphasize is the value of this diagnostic suite in experiments of this type, in particular the application of the radiography reconstruction routine to TNSA proton radiographs. Often, laser experiments which utilize proton radiography use a monoenergetic, imploding capsule-type backlighter; in such an experiment, the spatial resolution of the proton radiography (and therefore the structure of the reconstructed fields and currents) is inherently limited by the finite size of the backlighter, while TNSA radiography has the significant advantage of a considerably smaller proton source size. By combining the TNSA radiography with the reconstruction algorithm, we are able to utilize both to their fullest potential to examine the evolution of the experiment on both rapid timescales and small spatial scales.
		
		Thus, these experiments provide the first evidence of plasmoid formation in high-$\beta$ magnetic reconnection. The spatial structure and temporal evolution of proton radiographs reveal structure of primary islands, disclosing processes in plasmoid-dominated reconnection. These experiments provide evidence potentially validating aspects of current plasmoid theories and simulations, and the physical basis for explaining the observed fast magnetic reconnections in astrophysics and laboratory experiments. This work not only advances our knowledge of observed fast magnetic reconnection, but also opens up many novel opportunities along this line to solidify our understanding of exciting astrophysical and laboratory phenomena.

		\begin{acknowledgments}
			This work was supported in part by US DOE/NNSA CoE contract DE-NA0003868, US DoE/NNSA Award No. DE-NA0003856, NLUF contract DE-NA0003938, US DOE (Grant No. DE-FG03-09NA29553, No.DE-SC0007168), LLE (No.414090-G), NSF-DOE Partnership in Basic Plasma Science and Engineering award no. PHY-2010136, and joint NSF/DOE-NNSA award no. PHY2108050.
		\end{acknowledgments}

		\bibliography{mybib2}

\begin{thebibliography}{37}%
\makeatletter
\providecommand \@ifxundefined [1]{%
 \@ifx{#1\undefined}
}%
\providecommand \@ifnum [1]{%
 \ifnum #1\expandafter \@firstoftwo
 \else \expandafter \@secondoftwo
 \fi
}%
\providecommand \@ifx [1]{%
 \ifx #1\expandafter \@firstoftwo
 \else \expandafter \@secondoftwo
 \fi
}%
\providecommand \natexlab [1]{#1}%
\providecommand \enquote  [1]{``#1''}%
\providecommand \bibnamefont  [1]{#1}%
\providecommand \bibfnamefont [1]{#1}%
\providecommand \citenamefont [1]{#1}%
\providecommand \href@noop [0]{\@secondoftwo}%
\providecommand \href [0]{\begingroup \@sanitize@url \@href}%
\providecommand \@href[1]{\@@startlink{#1}\@@href}%
\providecommand \@@href[1]{\endgroup#1\@@endlink}%
\providecommand \@sanitize@url [0]{\catcode `\\12\catcode `\$12\catcode
  `\&12\catcode `\#12\catcode `\^12\catcode `\_12\catcode `\%12\relax}%
\providecommand \@@startlink[1]{}%
\providecommand \@@endlink[0]{}%
\providecommand \url  [0]{\begingroup\@sanitize@url \@url }%
\providecommand \@url [1]{\endgroup\@href {#1}{\urlprefix }}%
\providecommand \urlprefix  [0]{URL }%
\providecommand \Eprint [0]{\href }%
\providecommand \doibase [0]{http://dx.doi.org/}%
\providecommand \selectlanguage [0]{\@gobble}%
\providecommand \bibinfo  [0]{\@secondoftwo}%
\providecommand \bibfield  [0]{\@secondoftwo}%
\providecommand \translation [1]{[#1]}%
\providecommand \BibitemOpen [0]{}%
\providecommand \bibitemStop [0]{}%
\providecommand \bibitemNoStop [0]{.\EOS\space}%
\providecommand \EOS [0]{\spacefactor3000\relax}%
\providecommand \BibitemShut  [1]{\csname bibitem#1\endcsname}%
\let\auto@bib@innerbib\@empty
\bibitem [{\citenamefont {Biskamp}(2000)}]{cit1}%
  \BibitemOpen
  \bibfield  {author} {\bibinfo {author} {\bibfnamefont {D.}~\bibnamefont
  {Biskamp}},\ }\href@noop {} {\emph {\bibinfo {title} {Magnetic {Reconnection}
  in {Plasmas}}}}\ (\bibinfo  {publisher} {Cambridge University Press},\
  \bibinfo {address} {Cambridge},\ \bibinfo {year} {2000})\ \bibinfo {note}
  {oCLC: 668201161}\BibitemShut {NoStop}%
\bibitem [{\citenamefont {Yamada}\ \emph {et~al.}(2010)\citenamefont {Yamada},
  \citenamefont {Kulsrud},\ and\ \citenamefont {Ji}}]{cit2}%
  \BibitemOpen
  \bibfield  {author} {\bibinfo {author} {\bibfnamefont {M.}~\bibnamefont
  {Yamada}}, \bibinfo {author} {\bibfnamefont {R.}~\bibnamefont {Kulsrud}}, \
  and\ \bibinfo {author} {\bibfnamefont {H.}~\bibnamefont {Ji}},\ }\href
  {\doibase 10.1103/RevModPhys.82.603} {\bibfield  {journal} {\bibinfo
  {journal} {Reviews of Modern Physics}\ }\textbf {\bibinfo {volume} {82}},\
  \bibinfo {pages} {603} (\bibinfo {year} {2010})}\BibitemShut {NoStop}%
\bibitem [{\citenamefont {Zweibel}\ and\ \citenamefont {Yamada}(2009)}]{cit3}%
  \BibitemOpen
  \bibfield  {author} {\bibinfo {author} {\bibfnamefont {E.~G.}\ \bibnamefont
  {Zweibel}}\ and\ \bibinfo {author} {\bibfnamefont {M.}~\bibnamefont
  {Yamada}},\ }\href {\doibase 10.1146/annurev-astro-082708-101726} {\bibfield
  {journal} {\bibinfo  {journal} {Annual Review of Astronomy and Astrophysics}\
  }\textbf {\bibinfo {volume} {47}},\ \bibinfo {pages} {291} (\bibinfo {year}
  {2009})}\BibitemShut {NoStop}%
\bibitem [{\citenamefont {Masuda}\ \emph {et~al.}(1994)\citenamefont {Masuda},
  \citenamefont {Kosugi}, \citenamefont {Hara} \emph {et~al.}}]{cit4}%
  \BibitemOpen
  \bibfield  {author} {\bibinfo {author} {\bibfnamefont {S.}~\bibnamefont
  {Masuda}}, \bibinfo {author} {\bibfnamefont {T.}~\bibnamefont {Kosugi}},
  \bibinfo {author} {\bibfnamefont {H.}~\bibnamefont {Hara}},  \emph {et~al.},\
  }\href {\doibase 10.1038/371495a0} {\bibfield  {journal} {\bibinfo  {journal}
  {Nature}\ }\textbf {\bibinfo {volume} {371}},\ \bibinfo {pages} {495}
  (\bibinfo {year} {1994})}\BibitemShut {NoStop}%
\bibitem [{\citenamefont {Phan}\ \emph {et~al.}(2006)\citenamefont {Phan},
  \citenamefont {Gosling}, \citenamefont {Davis} \emph {et~al.}}]{cit5}%
  \BibitemOpen
  \bibfield  {author} {\bibinfo {author} {\bibfnamefont {T.~D.}\ \bibnamefont
  {Phan}}, \bibinfo {author} {\bibfnamefont {J.~T.}\ \bibnamefont {Gosling}},
  \bibinfo {author} {\bibfnamefont {M.~S.}\ \bibnamefont {Davis}},  \emph
  {et~al.},\ }\href {\doibase 10.1038/nature04393} {\bibfield  {journal}
  {\bibinfo  {journal} {Nature}\ }\textbf {\bibinfo {volume} {439}},\ \bibinfo
  {pages} {175} (\bibinfo {year} {2006})}\BibitemShut {NoStop}%
\bibitem [{\citenamefont {Taylor}(1986)}]{cit6}%
  \BibitemOpen
  \bibfield  {author} {\bibinfo {author} {\bibfnamefont {J.~B.}\ \bibnamefont
  {Taylor}},\ }\href {\doibase 10.1103/RevModPhys.58.741} {\bibfield  {journal}
  {\bibinfo  {journal} {Reviews of Modern Physics}\ }\textbf {\bibinfo {volume}
  {58}},\ \bibinfo {pages} {741} (\bibinfo {year} {1986})}\BibitemShut
  {NoStop}%
\bibitem [{\citenamefont {Olson}\ \emph {et~al.}(2016)\citenamefont {Olson},
  \citenamefont {Egedal}, \citenamefont {Greess} \emph {et~al.}}]{cit7}%
  \BibitemOpen
  \bibfield  {author} {\bibinfo {author} {\bibfnamefont {J.}~\bibnamefont
  {Olson}}, \bibinfo {author} {\bibfnamefont {J.}~\bibnamefont {Egedal}},
  \bibinfo {author} {\bibfnamefont {S.}~\bibnamefont {Greess}},  \emph
  {et~al.},\ }\href {\doibase 10.1103/PhysRevLett.116.255001} {\bibfield
  {journal} {\bibinfo  {journal} {Physical Review Letters}\ }\textbf {\bibinfo
  {volume} {116}},\ \bibinfo {pages} {255001} (\bibinfo {year}
  {2016})}\BibitemShut {NoStop}%
\bibitem [{\citenamefont {Jara-Almonte}\ \emph {et~al.}(2016)\citenamefont
  {Jara-Almonte}, \citenamefont {Ji}, \citenamefont {Yamada} \emph
  {et~al.}}]{cit8}%
  \BibitemOpen
  \bibfield  {author} {\bibinfo {author} {\bibfnamefont {J.}~\bibnamefont
  {Jara-Almonte}}, \bibinfo {author} {\bibfnamefont {H.}~\bibnamefont {Ji}},
  \bibinfo {author} {\bibfnamefont {M.}~\bibnamefont {Yamada}},  \emph
  {et~al.},\ }\href {\doibase 10.1103/PhysRevLett.117.095001} {\bibfield
  {journal} {\bibinfo  {journal} {Physical Review Letters}\ }\textbf {\bibinfo
  {volume} {117}},\ \bibinfo {pages} {095001} (\bibinfo {year}
  {2016})}\BibitemShut {NoStop}%
\bibitem [{\citenamefont {Hare}\ \emph {et~al.}(2017)\citenamefont {Hare},
  \citenamefont {Suttle}, \citenamefont {Lebedev} \emph {et~al.}}]{cit9}%
  \BibitemOpen
  \bibfield  {author} {\bibinfo {author} {\bibfnamefont {J.}~\bibnamefont
  {Hare}}, \bibinfo {author} {\bibfnamefont {L.}~\bibnamefont {Suttle}},
  \bibinfo {author} {\bibfnamefont {S.}~\bibnamefont {Lebedev}},  \emph
  {et~al.},\ }\href {\doibase 10.1103/PhysRevLett.118.085001} {\bibfield
  {journal} {\bibinfo  {journal} {Physical Review Letters}\ }\textbf {\bibinfo
  {volume} {118}},\ \bibinfo {pages} {085001} (\bibinfo {year}
  {2017})}\BibitemShut {NoStop}%
\bibitem [{\citenamefont {Nilson}\ \emph {et~al.}(2006)\citenamefont {Nilson},
  \citenamefont {Willingale}, \citenamefont {Kaluza} \emph {et~al.}}]{cit10}%
  \BibitemOpen
  \bibfield  {author} {\bibinfo {author} {\bibfnamefont {P.~M.}\ \bibnamefont
  {Nilson}}, \bibinfo {author} {\bibfnamefont {L.}~\bibnamefont {Willingale}},
  \bibinfo {author} {\bibfnamefont {M.~C.}\ \bibnamefont {Kaluza}},  \emph
  {et~al.},\ }\href {\doibase 10.1103/PhysRevLett.97.255001} {\bibfield
  {journal} {\bibinfo  {journal} {Physical Review Letters}\ }\textbf {\bibinfo
  {volume} {97}},\ \bibinfo {pages} {255001} (\bibinfo {year}
  {2006})}\BibitemShut {NoStop}%
\bibitem [{\citenamefont {Li}\ \emph {et~al.}(2007)\citenamefont {Li},
  \citenamefont {Séguin}, \citenamefont {Frenje} \emph {et~al.}}]{cit11}%
  \BibitemOpen
  \bibfield  {author} {\bibinfo {author} {\bibfnamefont {C.~K.}\ \bibnamefont
  {Li}}, \bibinfo {author} {\bibfnamefont {F.~H.}\ \bibnamefont {Séguin}},
  \bibinfo {author} {\bibfnamefont {J.~A.}\ \bibnamefont {Frenje}},  \emph
  {et~al.},\ }\href {\doibase 10.1103/PhysRevLett.99.055001} {\bibfield
  {journal} {\bibinfo  {journal} {Physical Review Letters}\ }\textbf {\bibinfo
  {volume} {99}},\ \bibinfo {pages} {055001} (\bibinfo {year}
  {2007})}\BibitemShut {NoStop}%
\bibitem [{\citenamefont {B}(2016)}]{cit12}%
  \BibitemOpen
  \bibinfo {editor} {\bibfnamefont {Lenhert}~\bibnamefont {B}},\ ed.,\ \href@noop {}
  {\emph {\bibinfo {title} {Electromagnetic phenomena in cosmical physics}}}\
  (\bibinfo {year} {2016})\ \bibinfo {note} {oCLC: 936532136}\BibitemShut
  {NoStop}%
\bibitem [{\citenamefont {Parker}(1957)}]{cit13}%
  \BibitemOpen
  \bibfield  {author} {\bibinfo {author} {\bibfnamefont {E.~N.}\ \bibnamefont
  {Parker}},\ }\href {\doibase 10.1029/JZ062i004p00509} {\bibfield  {journal}
  {\bibinfo  {journal} {Journal of Geophysical Research}\ }\textbf {\bibinfo
  {volume} {62}},\ \bibinfo {pages} {509} (\bibinfo {year} {1957})}\BibitemShut
  {NoStop}%
\bibitem [{\citenamefont {Loureiro}\ \emph {et~al.}(2007)\citenamefont
  {Loureiro}, \citenamefont {Schekochihin},\ and\ \citenamefont
  {Cowley}}]{cit14}%
  \BibitemOpen
  \bibfield  {author} {\bibinfo {author} {\bibfnamefont {N.~F.}\ \bibnamefont
  {Loureiro}}, \bibinfo {author} {\bibfnamefont {A.~A.}\ \bibnamefont
  {Schekochihin}}, \ and\ \bibinfo {author} {\bibfnamefont {S.~C.}\
  \bibnamefont {Cowley}},\ }\href {\doibase 10.1063/1.2783986} {\bibfield
  {journal} {\bibinfo  {journal} {Physics of Plasmas}\ }\textbf {\bibinfo
  {volume} {14}},\ \bibinfo {pages} {100703} (\bibinfo {year}
  {2007})}\BibitemShut {NoStop}%
\bibitem [{\citenamefont {Bhattacharjee}\ \emph {et~al.}(2009)\citenamefont
  {Bhattacharjee}, \citenamefont {Huang}, \citenamefont {Yang},\ and\
  \citenamefont {Rogers}}]{cit15}%
  \BibitemOpen
  \bibfield  {author} {\bibinfo {author} {\bibfnamefont {A.}~\bibnamefont
  {Bhattacharjee}}, \bibinfo {author} {\bibfnamefont {Y.-M.}\ \bibnamefont
  {Huang}}, \bibinfo {author} {\bibfnamefont {H.}~\bibnamefont {Yang}}, \ and\
  \bibinfo {author} {\bibfnamefont {B.}~\bibnamefont {Rogers}},\ }\href
  {\doibase 10.1063/1.3264103} {\bibfield  {journal} {\bibinfo  {journal}
  {Physics of Plasmas}\ }\textbf {\bibinfo {volume} {16}},\ \bibinfo {pages}
  {112102} (\bibinfo {year} {2009})}\BibitemShut {NoStop}%
\bibitem [{\citenamefont {Daughton}\ \emph {et~al.}(2009)\citenamefont
  {Daughton}, \citenamefont {Roytershteyn}, \citenamefont {Albright} \emph
  {et~al.}}]{cit16}%
  \BibitemOpen
  \bibfield  {author} {\bibinfo {author} {\bibfnamefont {W.}~\bibnamefont
  {Daughton}}, \bibinfo {author} {\bibfnamefont {V.}~\bibnamefont
  {Roytershteyn}}, \bibinfo {author} {\bibfnamefont {B.~J.}\ \bibnamefont
  {Albright}},  \emph {et~al.},\ }\href {\doibase
  10.1103/PhysRevLett.103.065004} {\bibfield  {journal} {\bibinfo  {journal}
  {Physical Review Letters}\ }\textbf {\bibinfo {volume} {103}},\ \bibinfo
  {pages} {065004} (\bibinfo {year} {2009})}\BibitemShut {NoStop}%
\bibitem [{\citenamefont {Loureiro}\ and\ \citenamefont
  {Uzdensky}(2016)}]{cit17}%
  \BibitemOpen
  \bibfield  {author} {\bibinfo {author} {\bibfnamefont {N.~F.}\ \bibnamefont
  {Loureiro}}\ and\ \bibinfo {author} {\bibfnamefont {D.~A.}\ \bibnamefont
  {Uzdensky}},\ }\href {\doibase 10.1088/0741-3335/58/1/014021} {\bibfield
  {journal} {\bibinfo  {journal} {Plasma Physics and Controlled Fusion}\
  }\textbf {\bibinfo {volume} {58}},\ \bibinfo {pages} {014021} (\bibinfo
  {year} {2016})}\BibitemShut {NoStop}%
\bibitem [{\citenamefont {Baalrud}\ \emph {et~al.}(2011)\citenamefont
  {Baalrud}, \citenamefont {Bhattacharjee}, \citenamefont {Huang} \emph
  {et~al.}}]{cit18}%
  \BibitemOpen
  \bibfield  {author} {\bibinfo {author} {\bibfnamefont {S.~D.}\ \bibnamefont
  {Baalrud}}, \bibinfo {author} {\bibfnamefont {A.}~\bibnamefont
  {Bhattacharjee}}, \bibinfo {author} {\bibfnamefont {Y.-M.}\ \bibnamefont
  {Huang}},  \emph {et~al.},\ }\href {\doibase 10.1063/1.3633473} {\bibfield
  {journal} {\bibinfo  {journal} {Physics of Plasmas}\ }\textbf {\bibinfo
  {volume} {18}},\ \bibinfo {pages} {092108} (\bibinfo {year}
  {2011})}\BibitemShut {NoStop}%
\bibitem [{\citenamefont {Ji}\ and\ \citenamefont {Daughton}(2011)}]{cit19}%
  \BibitemOpen
  \bibfield  {author} {\bibinfo {author} {\bibfnamefont {H.}~\bibnamefont
  {Ji}}\ and\ \bibinfo {author} {\bibfnamefont {W.}~\bibnamefont {Daughton}},\
  }\href {\doibase 10.1063/1.3647505} {\bibfield  {journal} {\bibinfo
  {journal} {Physics of Plasmas}\ }\textbf {\bibinfo {volume} {18}},\ \bibinfo
  {pages} {111207} (\bibinfo {year} {2011})}\BibitemShut {NoStop}%
\bibitem [{\citenamefont {Uzdensky}\ \emph {et~al.}(2010)\citenamefont
  {Uzdensky}, \citenamefont {Loureiro},\ and\ \citenamefont
  {Schekochihin}}]{cit20}%
  \BibitemOpen
  \bibfield  {author} {\bibinfo {author} {\bibfnamefont {D.~A.}\ \bibnamefont
  {Uzdensky}}, \bibinfo {author} {\bibfnamefont {N.~F.}\ \bibnamefont
  {Loureiro}}, \ and\ \bibinfo {author} {\bibfnamefont {A.~A.}\ \bibnamefont
  {Schekochihin}},\ }\href {\doibase 10.1103/PhysRevLett.105.235002} {\bibfield
   {journal} {\bibinfo  {journal} {Physical Review Letters}\ }\textbf {\bibinfo
  {volume} {105}},\ \bibinfo {pages} {235002} (\bibinfo {year}
  {2010})}\BibitemShut {NoStop}%
\bibitem [{\citenamefont {Hare}\ \emph {et~al.}(2018)\citenamefont {Hare},
  \citenamefont {Suttle}, \citenamefont {Lebedev} \emph {et~al.}}]{cit21}%
  \BibitemOpen
  \bibfield  {author} {\bibinfo {author} {\bibfnamefont {J.~D.}\ \bibnamefont
  {Hare}}, \bibinfo {author} {\bibfnamefont {L.~G.}\ \bibnamefont {Suttle}},
  \bibinfo {author} {\bibfnamefont {S.~V.}\ \bibnamefont {Lebedev}},  \emph
  {et~al.},\ }\href {\doibase 10.1063/1.5016280} {\bibfield  {journal}
  {\bibinfo  {journal} {Physics of Plasmas}\ }\textbf {\bibinfo {volume}
  {25}},\ \bibinfo {pages} {055703} (\bibinfo {year} {2018})}\BibitemShut
  {NoStop}%
\bibitem [{\citenamefont {Zylstra}\ \emph {et~al.}(2012)\citenamefont
  {Zylstra}, \citenamefont {Li}, \citenamefont {Rinderknecht} \emph
  {et~al.}}]{cit23}%
  \BibitemOpen
  \bibfield  {author} {\bibinfo {author} {\bibfnamefont {A.~B.}\ \bibnamefont
  {Zylstra}}, \bibinfo {author} {\bibfnamefont {C.~K.}\ \bibnamefont {Li}},
  \bibinfo {author} {\bibfnamefont {H.~G.}\ \bibnamefont {Rinderknecht}},
  \emph {et~al.},\ }\href {\doibase 10.1063/1.3680110} {\bibfield  {journal}
  {\bibinfo  {journal} {Review of Scientific Instruments}\ }\textbf {\bibinfo
  {volume} {83}},\ \bibinfo {pages} {013511} (\bibinfo {year}
  {2012})}\BibitemShut {NoStop}%
\bibitem [{\citenamefont {Bott}\ \emph {et~al.}(2017)\citenamefont {Bott},
  \citenamefont {Graziani}, \citenamefont {Tzeferacos} \emph {et~al.}}]{cit24}%
  \BibitemOpen
  \bibfield  {author} {\bibinfo {author} {\bibfnamefont {A.~F.~A.}\
  \bibnamefont {Bott}}, \bibinfo {author} {\bibfnamefont {C.}~\bibnamefont
  {Graziani}}, \bibinfo {author} {\bibfnamefont {P.}~\bibnamefont
  {Tzeferacos}},  \emph {et~al.},\ }\href {\doibase 10.1017/S0022377817000939}
  {\bibfield  {journal} {\bibinfo  {journal} {Journal of Plasma Physics}\
  }\textbf {\bibinfo {volume} {83}},\ \bibinfo {pages} {905830614} (\bibinfo
  {year} {2017})}\BibitemShut {NoStop}%
\bibitem [{\citenamefont {Froula}(2011)}]{cit25}%
  \BibitemOpen
  \bibinfo {editor} {\bibfnamefont {D.}~\bibnamefont {Froula}},\ ed.,\
  \href@noop {} {\emph {\bibinfo {title} {Plasma scattering of electromagnetic
  radiation: experiment, theory and computation}}},\ \bibinfo {edition} {1st}\
  ed.\ (\bibinfo  {publisher} {Elsevier},\ \bibinfo {address} {Amsterdam ;
  Boston},\ \bibinfo {year} {2011})\BibitemShut {NoStop}%
\bibitem [{\citenamefont {Fox}\ \emph {et~al.}(2020)\citenamefont {Fox},
  \citenamefont {Schaeffer}, \citenamefont {Rosenberg} \emph {et~al.}}]{cit22}%
  \BibitemOpen
  \bibfield  {author} {\bibinfo {author} {\bibfnamefont {W.}~\bibnamefont
  {Fox}}, \bibinfo {author} {\bibfnamefont {D.~B.}\ \bibnamefont {Schaeffer}},
  \bibinfo {author} {\bibfnamefont {M.~J.}\ \bibnamefont {Rosenberg}},  \emph
  {et~al.},\ }\href {http://arxiv.org/abs/2003.06351} {\enquote {\bibinfo
  {title} {Fast magnetic reconnection in highly-extended current sheets at the
  {National} {Ignition} {Facility}},}\ } (\bibinfo {year} {2020}),\ \bibinfo
  {note} {number: arXiv:2003.06351 arXiv:2003.06351 [physics]}\BibitemShut
  {NoStop}%
\bibitem [{\citenamefont {Rosenberg}\ \emph
  {et~al.}(2015{\natexlab{a}})\citenamefont {Rosenberg}, \citenamefont {Li},
  \citenamefont {Fox} \emph {et~al.}}]{cit26}%
  \BibitemOpen
  \bibfield  {author} {\bibinfo {author} {\bibfnamefont {M.}~\bibnamefont
  {Rosenberg}}, \bibinfo {author} {\bibfnamefont {C.}~\bibnamefont {Li}},
  \bibinfo {author} {\bibfnamefont {W.}~\bibnamefont {Fox}},  \emph {et~al.},\
  }\href {\doibase 10.1038/ncomms7190} {\bibfield  {journal} {\bibinfo
  {journal} {Nature Communications}\ }\textbf {\bibinfo {volume} {6}},\
  \bibinfo {pages} {6190} (\bibinfo {year} {2015}{\natexlab{a}})}\BibitemShut
  {NoStop}%
\bibitem [{\citenamefont {Carilli}\ and\ \citenamefont {Taylor}(2002)}]{cit27}%
  \BibitemOpen
  \bibfield  {author} {\bibinfo {author} {\bibfnamefont {C.~L.}\ \bibnamefont
  {Carilli}}\ and\ \bibinfo {author} {\bibfnamefont {G.~B.}\ \bibnamefont
  {Taylor}},\ }\href {\doibase 10.1146/annurev.astro.40.060401.093852}
  {\bibfield  {journal} {\bibinfo  {journal} {Annual Review of Astronomy and
  Astrophysics}\ }\textbf {\bibinfo {volume} {40}},\ \bibinfo {pages} {319}
  (\bibinfo {year} {2002})}\BibitemShut {NoStop}%
\bibitem [{\citenamefont {Govoni}\ \emph {et~al.}(2017)\citenamefont {Govoni},
  \citenamefont {Murgia}, \citenamefont {Vacca} \emph {et~al.}}]{cit28}%
  \BibitemOpen
  \bibfield  {author} {\bibinfo {author} {\bibfnamefont {F.}~\bibnamefont
  {Govoni}}, \bibinfo {author} {\bibfnamefont {M.}~\bibnamefont {Murgia}},
  \bibinfo {author} {\bibfnamefont {V.}~\bibnamefont {Vacca}},  \emph
  {et~al.},\ }\href {\doibase 10.1051/0004-6361/201630349} {\bibfield
  {journal} {\bibinfo  {journal} {Astronomy \& Astrophysics}\ }\textbf
  {\bibinfo {volume} {603}},\ \bibinfo {pages} {A122} (\bibinfo {year}
  {2017})}\BibitemShut {NoStop}%
\bibitem [{\citenamefont {Alt}\ and\ \citenamefont {Kunz}(2019)}]{cit29}%
  \BibitemOpen
  \bibfield  {author} {\bibinfo {author} {\bibfnamefont {A.}~\bibnamefont
  {Alt}}\ and\ \bibinfo {author} {\bibfnamefont {M.~W.}\ \bibnamefont {Kunz}},\
  }\href {\doibase 10.1017/S0022377819000084} {\bibfield  {journal} {\bibinfo
  {journal} {Journal of Plasma Physics}\ }\textbf {\bibinfo {volume} {85}},\
  \bibinfo {pages} {175850102} (\bibinfo {year} {2019})}\BibitemShut {NoStop}%
\bibitem [{\citenamefont {Waxer}\ \emph {et~al.}(2005)\citenamefont {Waxer},
  \citenamefont {Maywar}, \citenamefont {Kelly} \emph {et~al.}}]{cit30}%
  \BibitemOpen
  \bibfield  {author} {\bibinfo {author} {\bibfnamefont {L.}~\bibnamefont
  {Waxer}}, \bibinfo {author} {\bibfnamefont {D.}~\bibnamefont {Maywar}},
  \bibinfo {author} {\bibfnamefont {J.}~\bibnamefont {Kelly}},  \emph
  {et~al.},\ }\href {\doibase 10.1364/OPN.16.7.000030} {\bibfield  {journal}
  {\bibinfo  {journal} {Optics and Photonics News}\ }\textbf {\bibinfo {volume}
  {16}},\ \bibinfo {pages} {30} (\bibinfo {year} {2005})}\BibitemShut {NoStop}%
\bibitem [{\citenamefont {Rosenberg}\ \emph
  {et~al.}(2015{\natexlab{b}})\citenamefont {Rosenberg}, \citenamefont {Li},
  \citenamefont {Fox} \emph {et~al.}}]{cit33}%
  \BibitemOpen
  \bibfield  {author} {\bibinfo {author} {\bibfnamefont {M.}~\bibnamefont
  {Rosenberg}}, \bibinfo {author} {\bibfnamefont {C. K.}~\bibnamefont {Li}},
  \bibinfo {author} {\bibfnamefont {W.}~\bibnamefont {Fox}},  \emph {et~al.},\
  }\href {\doibase 10.1103/PhysRevLett.114.205004} {\bibfield  {journal}
  {\bibinfo  {journal} {Physical Review Letters}\ }\textbf {\bibinfo {volume}
  {114}},\ \bibinfo {pages} {205004} (\bibinfo {year}
  {2015}{\natexlab{b}})}\BibitemShut {NoStop}%
\bibitem [{\citenamefont {Biermann}(1950)}]{cit31}%
  \BibitemOpen
  \bibfield  {author} {\bibinfo {author} {\bibfnamefont {L.}~\bibnamefont
  {Biermann}},\ }\href {https://ui.adsabs.harvard.edu/abs/1950ZNatA...5...65B}
  {\bibfield  {journal} {\bibinfo  {journal} {Zeitschrift Naturforschung Teil
  A}\ }\textbf {\bibinfo {volume} {5}},\ \bibinfo {pages} {65} (\bibinfo {year}
  {1950})},\ \bibinfo {note} {aDS Bibcode: 1950ZNatA...5...65B}\BibitemShut
  {NoStop}%
\bibitem [{\citenamefont {Hatchett}\ \emph {et~al.}(2000)\citenamefont
  {Hatchett}, \citenamefont {Brown}, \citenamefont {Cowan} \emph
  {et~al.}}]{cit32}%
  \BibitemOpen
  \bibfield  {author} {\bibinfo {author} {\bibfnamefont {S.~P.}\ \bibnamefont
  {Hatchett}}, \bibinfo {author} {\bibfnamefont {C.~G.}\ \bibnamefont {Brown}},
  \bibinfo {author} {\bibfnamefont {T.~E.}\ \bibnamefont {Cowan}},  \emph
  {et~al.},\ }\href {\doibase 10.1063/1.874030} {\bibfield  {journal} {\bibinfo
   {journal} {Physics of Plasmas}\ }\textbf {\bibinfo {volume} {7}},\ \bibinfo
  {pages} {2076} (\bibinfo {year} {2000})}\BibitemShut {NoStop}%
\bibitem [{\citenamefont {Li}\ \emph {et~al.}(2006)\citenamefont {Li},
  \citenamefont {Séguin}, \citenamefont {Frenje} \emph
  {et~al.}}]{double_bubble}%
  \BibitemOpen
  \bibfield  {author} {\bibinfo {author} {\bibfnamefont {C.~K.}\ \bibnamefont
  {Li}}, \bibinfo {author} {\bibfnamefont {F.~H.}\ \bibnamefont {Séguin}},
  \bibinfo {author} {\bibfnamefont {J.~A.}\ \bibnamefont {Frenje}},  \emph
  {et~al.},\ }\href {\doibase 10.1103/PhysRevLett.97.135003} {\bibfield
  {journal} {\bibinfo  {journal} {Physical Review Letters}\ }\textbf {\bibinfo
  {volume} {97}},\ \bibinfo {pages} {135003} (\bibinfo {year}
  {2006})}\BibitemShut {NoStop}%
\bibitem [{\citenamefont {Uzdensky}\ and\ \citenamefont
  {Loureiro}(2016)}]{cit34}%
  \BibitemOpen
  \bibfield  {author} {\bibinfo {author} {\bibfnamefont {D.}~\bibnamefont
  {Uzdensky}}\ and\ \bibinfo {author} {\bibfnamefont {N.}~\bibnamefont
  {Loureiro}},\ }\href {\doibase 10.1103/PhysRevLett.116.105003} {\bibfield
  {journal} {\bibinfo  {journal} {Physical Review Letters}\ }\textbf {\bibinfo
  {volume} {116}},\ \bibinfo {pages} {105003} (\bibinfo {year}
  {2016})}\BibitemShut {NoStop}%
\bibitem [{\citenamefont {Tolman}\ \emph {et~al.}(2018)\citenamefont {Tolman},
  \citenamefont {Loureiro},\ and\ \citenamefont {Uzdensky}}]{cit35}%
  \BibitemOpen
  \bibfield  {author} {\bibinfo {author} {\bibfnamefont {E.~A.}\ \bibnamefont
  {Tolman}}, \bibinfo {author} {\bibfnamefont {N.~F.}\ \bibnamefont
  {Loureiro}}, \ and\ \bibinfo {author} {\bibfnamefont {D.~A.}\ \bibnamefont
  {Uzdensky}},\ }\href {\doibase 10.1017/S002237781800017X} {\bibfield
  {journal} {\bibinfo  {journal} {Journal of Plasma Physics}\ }\textbf
  {\bibinfo {volume} {84}},\ \bibinfo {pages} {905840115} (\bibinfo {year}
  {2018})}\BibitemShut {NoStop}%
\bibitem [{\citenamefont {Drake}\ and\ \citenamefont {Lee}(1977)}]{cit36}%
  \BibitemOpen
  \bibfield  {author} {\bibinfo {author} {\bibfnamefont {J.~F.}\ \bibnamefont
  {Drake}}\ and\ \bibinfo {author} {\bibfnamefont {Y.~C.}\ \bibnamefont
  {Lee}},\ }\href {\doibase 10.1063/1.862017} {\bibfield  {journal} {\bibinfo
  {journal} {Physics of Fluids}\ }\textbf {\bibinfo {volume} {20}},\ \bibinfo
  {pages} {1341} (\bibinfo {year} {1977})}\BibitemShut {NoStop}%
\end{thebibliography}%
		
	\end{document}